\documentclass[aps,prb,twocolumn,groupedaddress,floatfix,showpacs]{revtex4-1}

\usepackage{epsfig}
\usepackage{amssymb}
\usepackage{amsmath}

\begin{document}

\title{Effective three-particle interactions in low-energy models for multiband systems}
\author{Stefan A. Maier and Carsten Honerkamp}
\affiliation{Institute for Theoretical Solid State Physics, RWTH Aachen University, D-52056 Aachen, Germany 
\\ and JARA - FIT Fundamentals of Future Information Technology}

\date{February 24, 2012}

\begin{abstract}
We discuss different approximations for effective low-energy interactions in multiband models for weakly correlated  electrons. In the study of Fermi surface instabilities of the conduction band(s), the standard approximation consists only keeping those terms in the bare interactions that couple only to the conduction band(s), while corrections due to virtual excitations into bands away from the Fermi surface are typically neglected. Here, using a functional renormalization group approach, we present an improved truncation for the treatment of the effective interactions in the conduction band that keeps track of the generated three-particle interactions (six-point term) and hence allows one to include important aspects of these virtual interband excitations. Within a simplified two-patch treatment of the conduction band, we demonstrate that these corrections can have a rather strong effect in parts of the phase diagram by changing the critical scales for various orderings and the phase boundaries.
\end{abstract}


\maketitle
\section{Introduction}
In various areas of recent condensed matter physics it has become clear that interaction effects in multiband systems can lead to interesting phenomena that usually do not occur in single-band models. An important example thereof is the gap structure of the iron pnictide superconductors, where calculations show that the gap structure of the superconducting state depends significantly on the intra- and interorbital interaction parameters, and on the orbital composition of the bands that are usually mixtures of several Fe $d$-orbitals (see, e.g., Refs.~\onlinecite{maier,kuroki,wang,platt1,platt2}). If this complexity was ignored, one would obtain rather isotropic gaps around the Fermi surfaces. The gap anisotropy arises  due to {\em orbital makeup}\cite{maier}, i.e.\ due to the additional wavevector dependence of the orbital components of the Bloch functions for the bands that give much more structure even to local interactions in these multi-orbital systems.

The additional structure in the multiband interactions also creates a stronger sensitivity to features and parameters of the respective effective low-energy model, compared to one-band models. This calls for a more comprehensive theoretical study of multi-orbital model systems. One important aspect is whether the orbital composition of the bands is altered by interaction or correlation effects beyond those already captured in density functional theory (DFT). Another aspect is that all these studies are performed within effective low-energy models where many bands outside a respective energy window have been absorbed into model parameters. The question that shall be in the focus of this work is if there are sizable additional corrections due to {\em virtual excitations in the bands outside the low-energy window} in which the effective model is formulated.  In particular we will try to estimate if these virtual excitation effects can be more important than orbital makeup effects. 

A concrete physical question that motivates such considerations occurs in the high-$T_c$ cuprates. Here a comparison of DFT electronic structure data and experimentally measured critical temperatures $T_c$ for superconductivity suggests\cite{pavarini} that lowering a so-called axial orbital in energy  from above toward the Fermi-surface-forming Cu 3$d_{x^2-y^2}$-orbital causes an increase of $T_c$, although the same change makes the Fermi surface more round and hence less favorable for antiferromagnetic spin fluctuations. In a recent study\cite{uebelacker} we have tried to assess if orbital makeup effects can cause such material trends, but found only very moderate improvement of $T_c$. Hence the questions is if the next step in a reduction of  approximations results additional increases of $T_c$. 

We already note at this stage that virtual-excitation corrections have been identified as an important mechanism to screen down local (Hubbard) interaction parameters. A currently popular scheme to capture this effect is the constrained-RPA approach\cite{aryasetiawan,imada} that sums up the RPA series for the Coulomb interaction with at least one intermediate line in the bubble being a high-energy excitation. In another way, the work described here can be understood as 
an attempt to partly include such contributions in a renormalization group approach for the effective low-energy model.

In order to detect superconducting instabilities and hence to estimate a superconducting energy scale theoretically, one should look for a divergence of the effective interactions at low scales. In one-loop RG studies this is accomplished by integrating out the single-particle modes of the effective model with a decreasing cutoff or RG scale (for a recent review, see Ref.~\onlinecite{metznerRMP}). In this procedure, one-loop corrections with both internal lines in the effective low-energy window are summed up. The initial two-particle interaction for this flow should include the influence of the excitations outside the low-energy window. We can try to estimate these corrections in second order perturbation theory, which gives rise to diagrams with two propagator lines connecting two interaction vertices.  For these diagrams we will have contributions with the two internal lines both in the high-energy range, and with one in the high-energy and the other in the low-energy window. The latter processes can be expected to be more important, as they usually come with the smaller energy denominator. In the following we will argue that systematically including these latter contributions into the solution of the effective low-energy problem requires either an approximate correction of the low-energy effective four-point vertex (two-particle interaction) or to improve the truncation so as to keep the effective six-point vertex (three-particle interaction) in the effective low-energy model. We will compare these two schemes quantitatively in a simplified model and demonstrate that the virtual excitation effects can indeed change critical scales for superconducting instabilities considerably.

\section{Model}
The basic model for this work is a two-orbital Hamiltonian where one of the two resulting band crosses the Fermi level, while the other one is separated by an energy gap. We will discuss several approximate ways to include the band away from the Fermi surface into effective one-band models for the band near the Fermi surface. 
The structure of the model we use is borrowed from an (effective) two-orbital model derived for high-$T_c$ cuprates, more precisely for the $ sp\sigma $ and $ dp \sigma $ orbitals from the local density approximation (LDA) band structure of 
$ {\rm YBa_2 Cu_3 O_7} $ \cite{andersen_jpcs_1995}, extended by inter- and intra-orbital interactions of strengths $ U $ and $ U' $,
\begin{align} \notag
 H  =& \sum_{{\bf k},\sigma} \left( \begin{array}{cc} f_{{\bf k},+,\sigma}^\dagger & f_{{\bf k},-, \sigma}^\dagger \end{array} \right)
  \left( \begin{array}{cc} A_{\bf k} & C_{\bf k} \\
			C_{\bf k} & B_{\bf k} \end{array} \right)
  \left( \begin{array}{c} f_{{\bf k},+,\sigma} \\ 
			f_{{\bf k},-,\sigma} \end{array} \right) \\ \label{eqn:hamiltonian}
 & + \frac{U}{2} \sum_{i,\alpha,\sigma} n_{i,\alpha,\sigma} n_{i,\alpha,-\sigma}
 + \frac{U'}{2} \sum_{i,\alpha,\sigma,\sigma'} n_{i,\alpha,\sigma} n_{i,-\alpha,\sigma'} \, .
\end{align}
Here, $ f_{{\bf k},\alpha,\sigma} $  and $ n_{i,\alpha,\sigma} = f_{i,\alpha,\sigma}^\dagger
 f_{i,\alpha,\sigma} $ denote the annihilation operator 
of an electron with momentum $ {\bf k} $, orbital $ \alpha $ and spin orientation $ \sigma $
and the density of such a fermion at site $ i $, respectively. 
The non-interacting part of the Hamiltonian is given by
\begin{align*}
 A_{\bf k} & = \Delta E + \left( 1 - u_{\bf k}^+ \right) w_s \, ,\\
 B_{\bf k} & =  \left( 1 - u_{\bf k}^+ \right) w_d \, ,\\
 C_{\bf k} & = - u_{\bf k}^- \sqrt{w_d w_s} \, ,\\
 u_{\bf k}^\pm & = \frac{1}{2} \left( \cos k_y \pm \cos k_x \right) \, ,
\end{align*}
 $ \Delta E $ being the band separation and $ w_s $ and $ w_d $ the widths of the $ s $- and $ d $-orbitals, respectively. Of course other interactions terms like a Hund- or non-local terms can be added, but this does not play any role for the considerations that follow.

Our motivation for studying this model is that serves as a simple test case for the development of an functional renormalization group (fRG) approach to multiband systems.
We do not aim at making predictions for a specific material that hold on a quantitative level.
Regarding $ {\rm YBa_2 Cu_3 O_7} $, our results should only be taken with a grain of salt.
The reason for this is twofold: First of all, a multi-orbital tight-binding model of the cuprates should include orbitals on the oxygen atoms. The two-orbital
 model given above, however, does not allow for the description of Varma currents\cite{varma_curr} and other types of intra-unit cell order\cite{fischer}.
Moreover, the fRG approach for fermions used in this work is only viable as long as the renormalized interaction stays weak, whereas, in the cuprates,
  realistic values for the bare interaction are already large compared to the bandwidth. Also other parameters will not always be chosen according to ab initio calculations in the following.

We now switch to an imaginary-time
 functional integral formalism with Grassmann fields $ \bar{\psi}_{k,\alpha,\sigma} $ and $ \psi_{k,\alpha,\sigma} $ corresponding to 
the operators $ f_{{\bf k},\alpha,\sigma}^\dagger $  and $ f_{{\bf k},\alpha,\sigma} $. These fields depend on the $ 1+2 $ momentum
$ k = (k_0,{\bf k}) $ with Matsubara frequency $ k_0 $.

\begin{figure}
 \includegraphics[width=8.5cm]{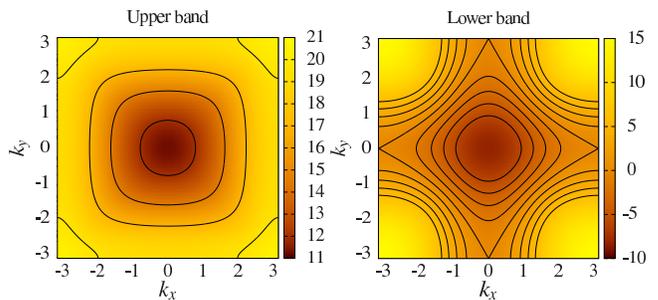}
 \caption{(Color online) Band structure of the two-band model for $ \Delta E =19.5 $, $ w_d=11.5 $ and $w_s=4.5 $ at van-Hove filling.  The lower band on the right is the conduction band. The colorbars on the sides indicate the band energy. This choice of parameters underlies the RG flow in Fig.~\ref{fig:nod-dSC}.}
\label{fig:bandstr}
\end{figure}

Diagonalization of the quadratic part $ S^{(2)} $ of the action corresponding to Eq.~(\ref{eqn:hamiltonian}) yields bands with energies 
\begin{equation*}
 E_{{\bf k},\pm} = \frac{1}{2} \left[ A_{\bf k} + B_{\bf k} \pm \sqrt{ \left( A_{\bf k} - B_{\bf k} \right)^2 + 4 C_{\bf k}^2} \right] \, ,
\end{equation*}
which are shown in Fig.~\ref{fig:bandstr}. We will work in a parameter range where the lower of the two bands cuts the Fermi level, while the upper band is entirely above the Fermi level. The effective low-energy model will the refer to a model with degrees of freedom in the functional integral that reside in the lower band near the Fermi surface only. This simple two-band set-up serves as minimal model for more complex situations in with many, possibly entangled bands in both high-energy and low-energy sector. 

The equation for the unitary transformation from orbitals $ \psi $ to bands $ \chi $ labeled by $\alpha = \pm$ reads as
\begin{align} \label{eqn:otb}
 \psi_{k,\alpha,\sigma} & = \alpha d_k \chi_{k,\alpha, \sigma} + c_k \chi_{k,-\alpha,\sigma} \, ,\\ \notag
 d_k & = \frac{N_{\bf k}}{2} \left( A_{\bf k} - B_{\bf k} + \sqrt{ \left( A_{\bf k} - B_{\bf k} \right)^2 + 4 C_{\bf k}^2 } \right) \, , \\ \notag
 c_k & = N_{\bf k} C_{\bf k} \, ,
\end{align}
where $ N_{\bf k} $ normalizes the transformation to a unitary one.
The inverse of this transformation gives the orbital amplitudes $\alpha d_k$ and $c_k$ for the band fields $\chi_{k,\alpha,\sigma}$. 
If we express the local interactions that are quartic in the fermion fields in terms of the band fields using Eq.~(\ref{eqn:otb}), we obtain four factors of $d_{k,\alpha,\sigma} $ or $c_{k,\alpha,\sigma} $ multiplying the interaction parameters  $U$ and $U'$. This collection of prefactors was dubbed {\em orbital makeup} \cite{maier} and obviously leads to some wavevector-dependence even of local interactions when they are expressed in band language.
In Appendix~\ref{sec:paramapp} we parameterize the interaction in the band picture in an efficient way, and later we will also discuss the impact of this orbital makeup on the critical scales for superconducting instabilities. 

\section{Effective low-energy action} \label{sec:efflowact}
Now we proceed towards an effective one-band action for the band near the Fermi level. This means we want to integrate out the upper band  
$E_{{\bf k},+}$ away from the Fermi level. In order to be concise, we use the matrix notation
\begin{equation*}
   \boldsymbol{ \bar{\phi} M \eta} = \sum_{k,\alpha,\sigma} \sum_{k',\alpha',\sigma'} 
 \bar{\phi}_{k,\alpha,\sigma} M_{k,\alpha,\sigma;k',\alpha',\sigma'} \eta_{k',\alpha',\sigma'} \, 
\end{equation*}
for the free part and other field-bilinears of the action. 

Consider now the generating functional $ W $ of the connected Green's functions
\begin{equation*}
 W [\bar{\eta}, \eta] = - \ln \left[ \int \! {\cal D} \chi \, e^{-S[\bar{\chi}, \chi] } 
e^{\bar{\boldsymbol \eta} {\boldsymbol \chi} +\bar{\boldsymbol \chi} {\boldsymbol \eta} } \right]
\end{equation*}
with source fields $ \eta $. We are interested in the low-energy properties of the system. 
{\em Before} taking derivatives with respect to the source fields, their upper-band components can thus be set to zero.
In the band language, the fields and the quadratic part $ S^{(2)} $ of the action can be decomposed  into lower and upper band parts
\begin{align*}
 S^{(2)} [\bar{\chi}, \chi] & = \boldsymbol{ \bar{\chi} D \chi} 
 = \boldsymbol{ \bar{\chi}}_+ \boldsymbol{D}_+ \boldsymbol{\chi}_+ + \boldsymbol{ \bar{\chi}}_- \boldsymbol{D}_- \boldsymbol{\chi}_-  \, , \end{align*} with \begin{align*}
 \boldsymbol{\chi} & = \boldsymbol{\chi}_+  + \boldsymbol{\chi}_-  \, , \quad  \boldsymbol{D} = \boldsymbol{D}_+  + \boldsymbol{D}_-  \, .
\end{align*}
Therefore, the covariance splitting formula \cite{salm_hon_2001} applies, which gives rise to the following form of the one-band effective action $ S_{\rm eff} $  for the lower band
\begin{align} \notag
 S_{\rm eff} [\bar{\chi}_-,\chi_-] 
& = \boldsymbol{\bar{\chi}}_- \boldsymbol{D}_- \boldsymbol{\chi}_- + {\cal V} [\bar{\chi_-},\chi_-] \, ,  \\ \label{eqn:Seff}
e^{-{\cal V} [\bar{\chi_-},\chi_-]} & =  \int \!\! {\cal D} \chi_+ \, e^{-\boldsymbol{\bar{\chi}}_+ \boldsymbol{D}_+ \boldsymbol{\chi}_+}
 e^{-S^{(4)} [\bar{\chi}_+ + \bar{\chi}_-, \chi_+ + \chi_-]} \, ,
\end{align}
which leads to
\begin{equation*}
 W [\bar{\eta}_-, \eta_-] = - \ln \left[ \int \! {\cal D} \chi_- \, e^{-S_{\rm eff} [\bar{\chi}_-, \chi_-] } 
e^{\bar{\boldsymbol \eta}_- {\boldsymbol \chi_-} +\bar{\boldsymbol \chi}_- {\boldsymbol \eta}_- } \right] \, .
\end{equation*}
The effective interaction $ {\cal V} $ contains a functional integral over the upper band part of the fields with measure $ {\cal D} \chi_+ $ and
 corresponds to the generating functional of amputated connected Greens functions, as used in the Polchinski renormalization group scheme \cite{Polchinski} (for a comprehensive reviews of the various generating functionals in our context, see e.g. Refs.~\onlinecite{salmhoferbook,enss_phd,kopietzbook}). This means that the parameters of the effective low-energy action are given by these amputated connected Green's functions.
In the special case of Eq.~(\ref{eqn:Seff}), however, only the upper band has been integrated out. Thus, in the diagrammatic expansion of the expansion coefficients in the fields $\chi_-$ and $\bar \chi_-$, the propagators on internal lines  are restricted to the 
upper band, whereas external legs live on the lower one.
Before we evaluate $ S_{\rm eff} $, we recall that the amputated connected Greens functions can be recovered from one-particle irreducible (1PI) diagrams by drawing all tree diagrams with 1PI vertices. In our case, the internal lines of these tree diagrams are high-energy propagators.

\begin{figure}
 \includegraphics[width=8.4cm]{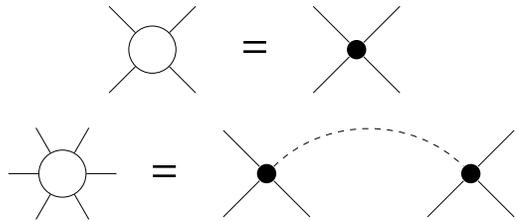}
 \caption{Four- and six-point vertex of $ S_{\rm eff} $ for the effective one-band model as considered at approximation level~2. Small filled vertices correspond to the bare interaction $ S^{(4)} $, whereas the vertices of 
$ S_{\rm eff} $ on the left hand sides are denoted by empty circles. Solid lines represent the lower and dashed lines the upper band. There are no propagators attached to the external legs. Self-energy effects will be neglected.}
\label{fig:6pt}
\end{figure}

For assessing the low energy properties of the two-band model, the general strategy is as follows.
In a first step, we calculate $ S_{\rm eff} $.
Since the Fermi surface does not intersect with the upper band, the diagrams  that appear in Eq.~(\ref{eqn:Seff}) need not be regularized and therefore the lower-band effective action 
can be evaluated perturbatively for sufficiently small values of the bare interaction.
In a second step, the effective action of the lower band is treated by a method of choice, which is in our case the fRG\cite{metznerRMP}. Then $ S_{\rm eff} $ imposes an initial condition on the RG flow. 

For our main focus, determining the energy scale for Cooper instabilities, the effective interaction of the low-energy model is the object of prime interest. The simplest truncation of the effective action would then be to drop all terms of order higher than four. 
Then the four-point term in the effective action, $ S_{\rm eff}^{(4)}$, can be computed in perturbation theory in the interactions involving high-energy modes.
In lowest order, i.e.\ zeroth order in the interactions with high-energy modes or first order in the bare interactions irrespective of energy scales, $ S_{\rm eff}^{(4)}$ is just the bare interaction of the fields in the lower band, decorated with orbital makeup. This is the standard that has been used e.g in all studies of unconventional pairing in the iron pnictides, or is used implicitly when the one-band Hubbard model is used as a model for the high-$T_c$ cuprates.  We will later denote this truncation as {\em approximation level~1}. Solving the low-energy theory diagrammatically captures (possibly singular) diagrams with both internal lines in the low-energy window.

In the next order for $S_{\rm eff}^{(4)}$, i.e.\ second order in the bare high-energy interactions, we get various diagrams. On one hand, there are self-energy Hartree- and Fock-like contributions on the external legs. Further, there are one-loop corrections with both lines in the high-energy sector. As mentioned, these are non-singular one-loop terms, as all internal lines are away from the Fermi level. Let us call the scheme that keeps these terms  {\em approximation 1'}. If we now again solve the low-energy theory diagrammatically, we will capture one-loop corrections for the effective interaction that have both propagator lines either in the high-energy range and that are already included in $S_{\rm eff}^{(4)}$, and corrections with both lines in the low-energy sector, coming from the perturbation expansion in $S_{\rm eff}^{(4)}$. What is not included are 'mixed' diagrams with one internal line in the high-energy range and one line in the low-energy window. Looking at the energy denominators of these lines, these excluded mixed contributions should be potentially more important than those with two internal high-energy propagators captured at approximation level~1'.

We could now go on and include further order in the bare interactions as corrections to $S_{\rm eff}^{(4)}$. It is however clear that in these higher-order diagrams all internal lines will be high-energy propagators. Hence, these corrections do not include the missing mixed diagrams, and thus we do not pursue these corrections any further here. In principle, they can be summed up using RG schemes, as described in Ref.~\onlinecite{kinza}.

The next useful extension would be to go along another path and to improve the truncation of the effective action. Hence we now keep the six-point term in the effective action. In the tree diagram expansion of $ S_{\rm eff} $ the sixth-order term is generated in second order in the 1PI four-point vertices of the high-energy theory. Here we will replace them by the lowest order, i.e.\ by the bare interactions, as shown in Fig.~\ref{fig:6pt}. This means that possible renormalizations of the two-particle interactions by additional high-energy processes are deliberately excluded. As argued above these corrections with additional high-energy propagators should however be smaller due to the energy separation of the bands.  This we will call {\em approximation level~2}.
Furthermore, we should potentially drop self-energy corrections in order to avoid double counting of contributions, i.~e.\ to the self-energy that are already included the DFT-based derivation in our two-orbital Hamiltonian. 
In this approximation, the quadratic part and the bare four-point couplings remain unrenormalized, whereas a six-point term depicted in Fig.~(\ref{fig:6pt}) is generated. For the effective action, we get
\begin{align*}
 S_{\rm eff}[\bar{\chi}_-,\chi_-]  = & \,  \left(S^{(2)} + S^{(4)} \right) [\bar{\chi}_-,\chi_-]  +S^{(6)}_{\rm eff} [\bar{\chi}_-,\chi_-] \\
 S^{(6)}_{\rm eff} [\bar{\chi}_-,\chi_-]  = & \, - \frac{1}{36} \int \! d \boldsymbol{\xi}  \, F^{(6)} (\boldsymbol{\xi} ) 
 \bar{\chi}_- (\xi_1) \bar{\chi}_- (\xi_2) \bar{\chi}_- (\xi_3) \\ & \times \chi_- (\xi_4) \chi_- (\xi_5) \chi_- (\xi_6) \, ,
\end{align*}
$ \xi_i = (k_i,\sigma_i) $ being a short-hand notation for the quantum numbers of the fields.
For the precise form of the six-point coupling function $ F^{(6)} $ we refer to Eq.~(\ref{eqn:6pt}) in the appendix.
Now, in a perturbative expansion for the low-energy theory, we will receive contributions where two legs of the six-point coupling will be folded together by a low-energy propagator line. As the six-point term came about by joining two legs of two four-point interactions by a high-energy propagator line, this will effectively bring in those missing diagrams with two internal propagators, one of which is a high-energy mode, and the other a low-energy mode.

We note in passing that the constrained RPA used for computing effective Hubbard interaction parameters\cite{aryasetiawan,imada} can be understood as an infinite order resummation of the mixed diagrams included at approximation level~2. Resummation at level~1', i.e.\ without any internal lines in the low-energy sector,  would presumably result in much less reduction of the onsite repulsion. On the other hand we note that just keeping the six-point term still does not capture the full cRPA series, as pure powers of mixed loops (i.e.\ bubbles with one high-energy and one low-energy line) included in the cRPA are not contained in the RPA series generated at our approximation level~2. This can be seen from constructing bubble sums with the elements of Fig.~\ref{fig:6pt} by contracting low-energy lines. There is a mixed diagram in second order in the bare interactions, but in third order or fourth order we have to add  a pure low-energy loop between two mixed loops. So, some orders in the mixed diagrams are still missing, but including the six-point term goes in the right direction. On the other hand, in contrast with the cRPA, our approximation does not neglect vertex corrections or particle-particle diagrams. The goal of the present study is to show the effects of these corrections to the simpler truncation. The question of how the cRPA series is understood in terms of the effective action will be discussed in another forthcoming publication.

\section{fRG treatment}
We are now in a position to proceed with the second step of solving the low-energy model, which we will do by a fRG flow. This will clarify the differences between the various levels of approximations.

\subsection{1PI functional RG scheme with smooth frequency cutoff}
We now introduce an infrared (IR) cutoff on the lower band by replacing $ {\boldsymbol D}_- $ by $ {\boldsymbol D}_- {\boldsymbol R}_\lambda^{-1} $, where $ {\boldsymbol R}_\lambda $ denotes a regulator function. In this paper, we choose
\begin{equation*}
 {\boldsymbol R}_\lambda (\xi,\xi') = \delta(\xi-\xi') \, \frac{k_0^2}{k_0^2+\lambda^2} \, .
\end{equation*}
\begin{figure}
 \includegraphics[width=8.5cm]{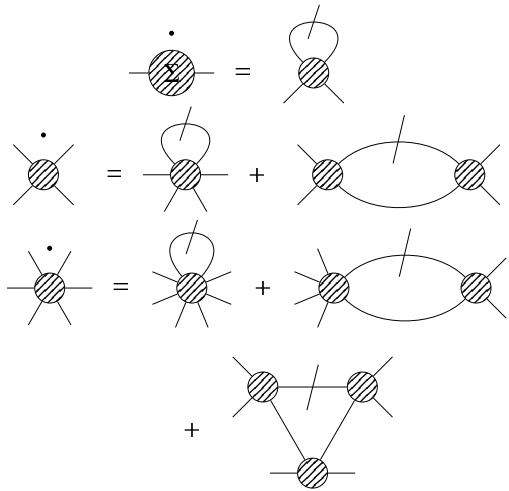}
 \caption{Flow equations for the self-energy, the four-point and six-point vertices, all one-particle-irreducible. A dot represents a derivative with respect to the cutoff.
 Lines with a slash correspond to the single-scale propagator $ {\boldsymbol S} = \dot{\boldsymbol G} - {\boldsymbol G} \dot{\boldsymbol \Sigma} {\boldsymbol G}$. For more details, see e.g. Ref.~\onlinecite{metznerRMP}.}
\label{fig:floweq}
\end{figure}

This particular choice of the regulator, introduced by Husemann and Salmhofer\cite{husemann_2009}, does not completely suppress contributions from the Fermi surface at nonzero $ \lambda $ and therefore allows one to 
take a possible ferromagnetic instability into account. 
Moreover, a pure frequency cutoff with ${\boldsymbol R}_\lambda =0$ at $k_0=0$  circumvents Fermi-surface-renormalization issues\cite{metznerRMP}, since the full propagator reads as
\begin{equation*}
 {\boldsymbol G} = {\boldsymbol R}_\lambda \left[ {\boldsymbol Q} + {\boldsymbol \Sigma} {\boldsymbol R}_\lambda \right]^{-1} \, ,
\end{equation*}
$ \boldsymbol \Sigma $ being the self-energy.
So self-energy effects may be neglected {\em without} ignoring the most relevant terms.

Starting from an exact flow equation for the generating functional $ \Gamma_- $ of the one-particle irreducible (1PI) vertex functions related to $ W [\bar{\eta}_-,
\eta_-] $ by a Legendre transformation in the lower-band fields, one obtains an infinite hierarchy of differential equations for the 1PI vertices \cite{salm_hon_2001,metznerRMP}. The subscript of $ \Gamma_- $ reminds us that only degrees of freedom in the lower band are integrated out, while the higher bands have been absorbed in the initial conditions.
The first three of the RG equations for the vertices generated by $\Gamma_-$ are expressed in a diagrammatic language in Fig.~\ref{fig:floweq}. 
At this point we note that there is no simple relation between $ \Gamma_- $ and the Legendre transform $ \Gamma $ of $ W [\bar{\eta},\eta] $. In the former case, 
information about correlations in the upper band is lost and the one-particle irreducibility only holds regarding propagators on the lower band. In the latter, namely the full multiband case, the band index however appears as an additional quantum number that is summed over on the internal lines of 1PI diagrams.

In order to make progress, the hierarchy of flow equations needs to be truncated at some point.
In the conventional truncation scheme one neglects the six-point vertex completely \cite{salm_hon_2001}. In the so-called Katanin scheme\cite{Katanin_trunc} six-point contributions that are generated during the RG flow are fed back into the flow equation for the four-point vertex. Both established truncation schemes however are not suited for a non-vanishing initial six-point vertex. So its impact on the flow poses a conceptually new problem.

\subsection{Improved truncation with one-loop six-point feedback} \label{subsec:imptrunc}
\begin{figure}
 \begin{center}
 \includegraphics[width=8.5cm]{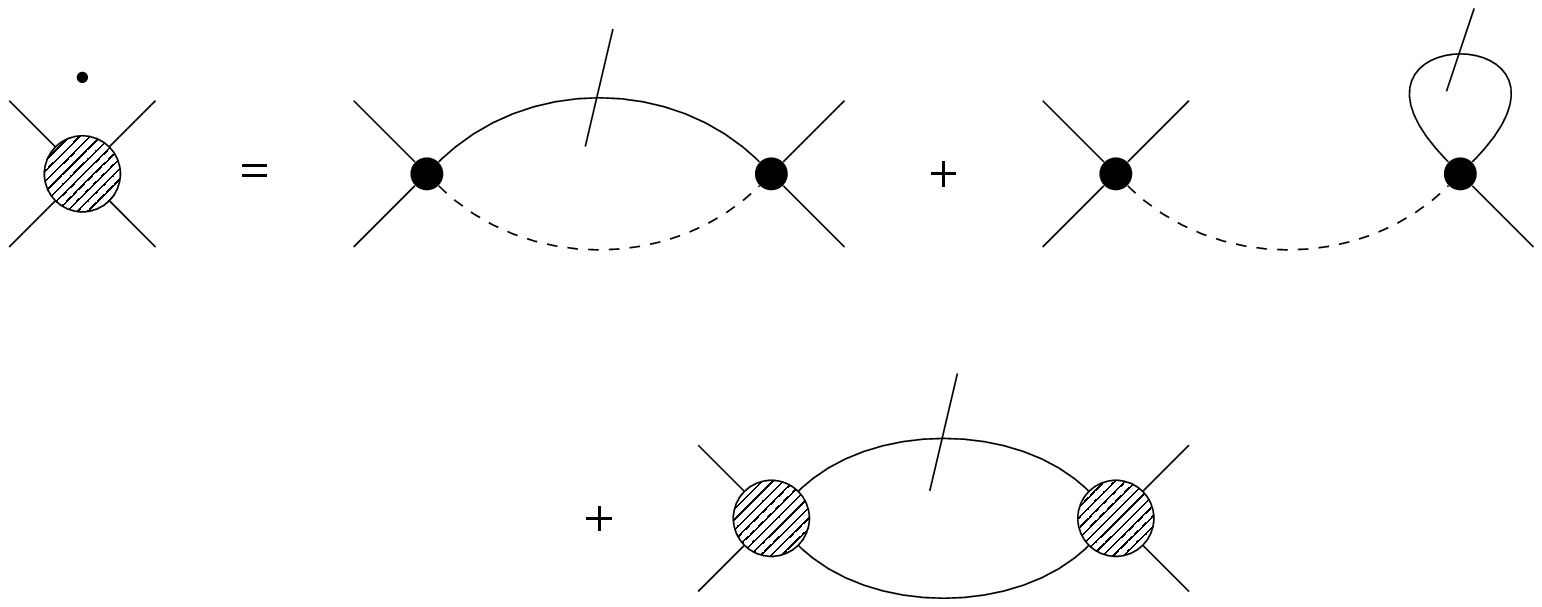}
 \end{center}
 \caption{Effective flow equation in the low-energy problem for the one-particle irreducible four-point vertex including the feedback term re-expressed in terms of the bare couplings. The dashed line stands for propagators in the upper band.}
\label{fig:feedback}
\end{figure}

Let us now return to the initial interactions of the low-energy problem in the conduction band that are given by the effective interactions after the  high-energy modes in the upper band have been integrated out.  As argued above, dropping all effective interactions higher than the four-point (two-particle) term ignores possibly important contributions. Keeping the six-point term in arbitrary order in the bare interactions or even keeping higher terms are hard tasks. A simpler, improved truncation scheme going beyond approximation level~1 consists in neglecting eight-point and higher interactions and assuming that the six-point vertex does not flow in the weak coupling regime. Then the six-point vertex would just be given by the product of two 1PI four-point vertices, connected by a high-energy propagator as depicted in the diagram in the lower part of Fig.~\ref{fig:6pt}.
 If the six-point vertex is then fed back into the four-point flow equation for the low-energy theory,
 the missing diagrams appear on the right-hand side with one high-energy and one single-scale low-energy line as shown in Fig.~\ref{fig:feedback}. 
 In this truncation, two-loop contributions are neglected.
Note that in Ref.~\onlinecite{Katanin-twoloop} also two-loop terms have been considered. However, this has only been done for the case of an initially vanishing six-point vertex and 
purely local bare interactions.

Before solving the flow equation in this truncation let us look briefly at the feedback term consisting of the first two diagrams in Fig.~\ref{fig:feedback}. Their precise form shall be given in Appendix~\ref{sec:6ptapp}. 
The fact that the diagrams with self-energy insertions in Fig.~\ref{fig:feedback} get disconnected when the line on the upper band is cut should not lead to confusion. Since we are dealing with an effective lower-band action, one-particle irreducibility only holds regarding lines on the lower band.
 If self-energy effects are neglected, the extra term in the flow equation for the four-point vertex is just identical to the scale derivative of the sum of second-order diagrams in the bare interaction that have one internal line one the upper band and another on the lower one. In the infrared, they stay regular but will be larger compared to second order diagrams with all internal lines on the upper band provided that the band separation is sufficiently large.
We therefore neglect those upper band diagrams and restrict the terms in the effective interactions to the tree level in the high-energy modes.

In the following, we shall distinguish three levels of approximation:

\begin{enumerate}
 \item Ignoring the six-point vertex completely. In this approximation the only multiband effects are the signatures of orbital makeup. We consider this conventional truncation for comparison in order to distinguish orbital makeup from six-point effects.
\item Including the feedback term in one-loop fRG and using the flow equation Fig.~\ref{fig:feedback}. 
\item Adding the mixed-band diagrams in the limit $ \lambda \to 0 $ to the initial condition for the flow of the low-energy model. This approximation should yield reliable results if the mixed diagrams are already close to their infrared value at scales at which the lower-band diagrams only have induced a small renormalization of the initial couplings. 
\end{enumerate}
Approximation levels 1 and 2 were already introduced in Sec.~\ref{sec:efflowact}, and level~3 is a simplification of level~2 that is easier to handle numerically. In the following we will compare the fRG flows in the low-energy models resulting from these approximations.

If one recalls that the LDA-derived dispersion of the underlying two-orbital model already contains interaction effects on a certain level, the band-flip self-energy insertion diagram (second term on the right hand side in the diagrammatic equation in Fig.~\ref{fig:feedback}) should potentially be neglected in order to avoid double counting. Its impact is however not important, as will be commented on at the end of Subsec.~\ref{sec:num6pt}.

\subsection{Two-patch approximation}
In order to make the resulting numerical calculations more feasible, and as we are mainly interested in getting a first picture of the effects due to the higher truncation, we now employ the so called two-patch approximation that has been used in the context of the one-band Hubbard model \cite{furukawa_1998,schulz} or iron pnictides\cite{chubukov}.

\begin{figure}
 \begin{center}
 \includegraphics[width=8.5cm]{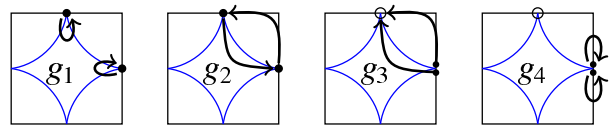}
 \end{center}
 \caption{The four running couplings $g_1$ to $g_4$ of the two-patch approximation for the conduction band. The blue lines represent the Fermi surface.}
\label{fig:gs}
\end{figure}

Just like the one-band Hubbard model the dispersion of the lower band of our model Hamiltonian in Eq.~(\ref{eqn:hamiltonian}) has saddle points at $ {\bf k} = A = (0,\pi) $ and $ B = (\pi,0) $. 
 If the system is now considered at van Hove filling, $ A $ and $ B $ lie on the Fermi surface.
 At zero temperature, the low energy properties then are dominated by contributions of a small vicinity around these saddle points. We therefore restrict the internal momenta in the lower band diagrams to two small patches around $ A $ and $ B $. In the following, we neglect self-energy effects and the frequency dependence of the coupling functions.  In the one-loop diagrams the momentum integral then only enters in the bare susceptibilities
\begin{align*}
 \Phi_{\rm pp} (l) & = \sum_{p_0} \int_{\rm patch} \!\!\!\!\!\! d{\bf p} \,  G_-(p) \, G_-(l-p) \\
 \Phi_{\rm ph} (l) & = \sum_{p_0} \int_{\rm patch} \!\!\!\!\!\! d{\bf p} \,  G_-(p) \, G_-(l+p) \, .
\end{align*}
For zero temperature, the Matsubara sum is evaluated analytically for $ l_0 =0 $ and the two-patch approximation restricts the transfer momenta $ {\bf l} $ to $ {\boldsymbol 0} $ or 
$ \hat{\boldsymbol \pi} = (\pi,\pi) $.
In our
truncation, the fRG analysis can now be restricted to four running couplings depicted in Fig.~\ref{fig:gs}, namely
\begin{align*}
 g_1 & = V_- (A,B,B) = V_-(B,A,A) \\
 g_2 & = V_- (A,B,A) = V_-(B,A,B) \\
 g_3 & = V_- (A,A,B) = V_-(B,B,A) \\
 g_4 & = V_- (A,A,A) = V_-(B,B,B) \, ,
\end{align*}
which drastically reduces the computational cost of the RG flow. At approximation level~1, the initial conditions for these couplings are obtained from transforming the intra- and interorbital interactions from the bare Hamiltonian into the band representation. The corresponding equations are given in Appendix \ref{sec:paramapp}, Eq.~(\ref{eqn:bare-coup-}).

As for the mixed-band diagrams, however, also regions away from the Fermi surface contribute significantly. Therefore, the loop integrals in the mixed-band diagrams are to be
taken over the whole Brillouin zone. 
In the case of the one-band Hubbard model, the
cutoff can be chosen such that the resulting loop integrals can be evaluated analytically and do not depend on the patch size \cite{Katanin-two-patch}. However, 
that cutoff scheme is only viable in a small neighborhood around the saddle points whereas the cutoff needs to be defined on the entire Brillouin zone in order to
consider the six-point feedback.
To the authors' knowledge, only momentum shell cutoff schemes have been used for the two-patch model while a frequency cutoff is used in this paper. 

The flow equations in the two-patch approximation read as
\begin{align} \notag
 \dot{g}_1  = & \, \delta  V_- (A,B,B) + d_1 \left( g_1^2 + g_3^2 \right) + 2 d_2 \left( g_2 - g_1 \right) g_4 \\ \label{eqn:gdot1}
& - d_3 \left( g_1^2 + g_2^2 \right) \\ \notag
 \dot{g}_2  = & \, \delta  V_- (A,B,A) + d_1 g_2 \left( g_1 - g_2 \right) + 2 d_2 g_2 g_4 \\ \label{eqn:gdot2}
& - 2 d_3  g_1 g_2 \\ \label{eqn:gdot3}
 \dot{g}_3  = & \, \delta  V_- (A,A,B) - 2 d_0 g_3 g_4 + 2 d_1 g_3 \left( 2 g_1 - g_2 \right) \\ \notag
 \dot{g}_4  = & \, \delta  V_- (A,A,A) - d_0 \left( g_3^2 + g_4^2 \right) \\ \label{eqn:gdot4}
 &+ d_2 \left( g_2^2 + 2 g_1 g_2 - 2 g_1^2 + g_4^2 \right) \, ,
\end{align}
where the dot denotes a derivative with respect to $ \lambda $. The six-point feedback leads to correction terms $ \delta V_- $ that do not occur in two-patch studies of one-band systems. They are given in appendix~\ref{sec:6ptapp}.  The integration over the patches in the loops
\begin{align*}
 d_0 & = \dot{\Phi}_{\rm pp} ({\boldsymbol 0}) \, , \quad d_1 =  \dot{\Phi}_{\rm ph} (\hat{\boldsymbol \pi}) \, , \\
 d_2 & = \dot{\Phi}_{\rm ph} ({\boldsymbol 0}) \, , \quad d_3 =  -\dot{\Phi}_{\rm pp} (\hat{\boldsymbol \pi})
\end{align*}
are performed numerically using an adaptive routine \cite{genz}.
The different levels of approximation introduced in Subsec. \ref{subsec:imptrunc} now imply the following:
\begin{enumerate}
 \item Neglecting $ \delta V_- $ and initializing the $ g_i $s by the respective values of the coupling function $ V_- $ at $ \lambda = \infty $.
 \item Keeping $ \delta V_- $ and initializing the $ g_i $s by the respective values of $ V_- $.
 \item Neglecting $ \delta V_- $ and initializing the $ g_i $s by the respective values of $ V_- + \Delta V_- $, where
$ \Delta V_- = - \int^\infty_0 d \lambda \, \delta V_- $ denotes the sum of all second-order mixed-band diagrams. 
\end{enumerate}
In all the examples considered in this paper, we observe an abrupt flow to strong coupling at some critical scale $ \lambda_{\rm crit} $.
In the two-patch approximation, the flow equations of the couplings to external source fields for $s$- and $ d $-wave 
superconductivity ($\alpha_{s{\rm SC}} $ and $ \alpha_{d{\rm SC}}$, respectively), anti-ferromagnetism ($\alpha_{\rm AF} $) and ferromagnetism ($\alpha_{\rm FM} $)
 take the simple forms
\begin{align*}
 \dot{\alpha}_{s{\rm SC}}  & = - 2 d_0 \, (g_3 + g_4) \, \alpha_{s {\rm SC}}  \\
 \dot{\alpha}_{d{\rm SC}}  & = - 2 d_0 \, (g_4 - g_3) \,\alpha_{d {\rm SC}}   \\
 \dot{\alpha}_{\rm AF}  & = + 2 d_1 \, (g_1 + g_3) \, \alpha_{\rm AF}  \\
 \dot{\alpha}_{\rm FM}  & = + 2 d_2 \, (g_2 + g_4) \, \alpha_{\rm FM}  \, .
\end{align*}
These couplings also appear in the expressions for the corresponding susceptibilities and determine which ordering tendency grows fastest at the critical
scale, i.e.\ is the leading instability at $ \lambda_{\rm crit} $.

\section{Numerical results} \label{sec:num}

We now solve the two-patch model at the three approximation levels described before for the parameters given in Tab.~\ref{tab:param} at van Hove filling. We first discuss the approximation level~1, i.e.\ without six-point feedback and then continue to describe the changes in the higher approximation levels. 

\subsection{Flows without six-point term: Impact of orbital makeup and competition of FM and $ d $SC instabilities} \label{sec:numwo}

In this subsection we discuss the impact of orbital makeup when the six-point feedback is neglected. 

Let us first consider the case of vanishing inter-orbital interaction, $U'=0$.
In that case, according to Eq.~(\ref{eqn:bare-coup-}) 
all four couplings take on the same value $ U_{\rm eff} $ at the beginning of the flow. 
 Since $ d^2_k + c^2_k =1 $, these couplings
are smaller than the intra-orbital coupling $ U $. Moreover, upon an expansion of the lower band dispersion around the saddle points,
 the lower-band dispersion is equivalent to the
dispersion of the one-band $ t$-$t'$ Hubbard model up to second order with effective parameters $ t $ and $ t' $ for next nearest and next-to-nearest neighbor hopping.
So for $ U' = 0 $ and approximation level~1, where the upper band does not enter, we are back to the one-band $ t $-$ t'$ model in a two-patch approximation, albeit with a reduced $U$ and employing a smooth frequency cutoff unlike the the momentum-shell cutoff in previous studies of this model. 

\begin{figure*}
 \includegraphics[width=17.8cm]{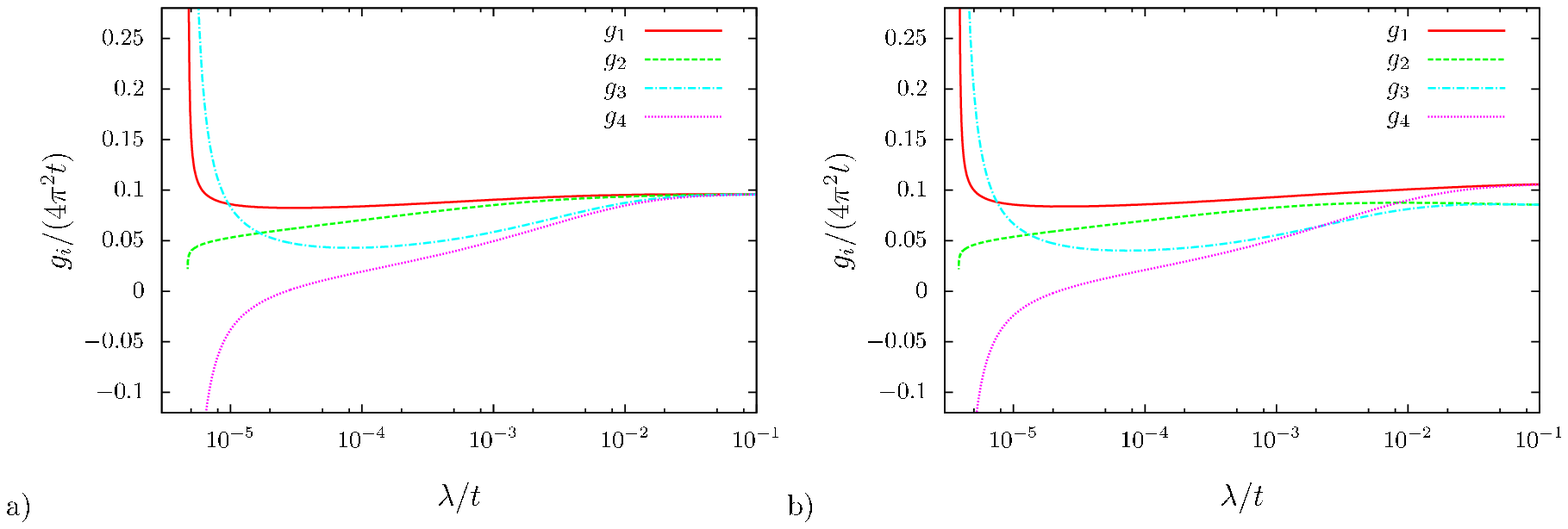}
 \caption{(Color online) Flow in the low-energy model for $ \Delta E =19.5 $, $ w_d=11.5 $, $w_s=4.5 $, $ U=0.03 \cdot 4 \pi^2 w_d $,
 $ U'=0 $ (a) and $ U'=0.3 U $ (b) at approximation level~1. The two-patch couplings are depicted as solid ($ g_1 $), dashed ($ g_2 $), dashed-dotted ($ g_3 $) and dotted lines ($ g_4 $).
The band-flip self-energy term in the feedback has not been taken into account.}
 \label{fig:nod-dSC}
\end{figure*}

\begin{figure}
 \includegraphics[width=8.5cm]{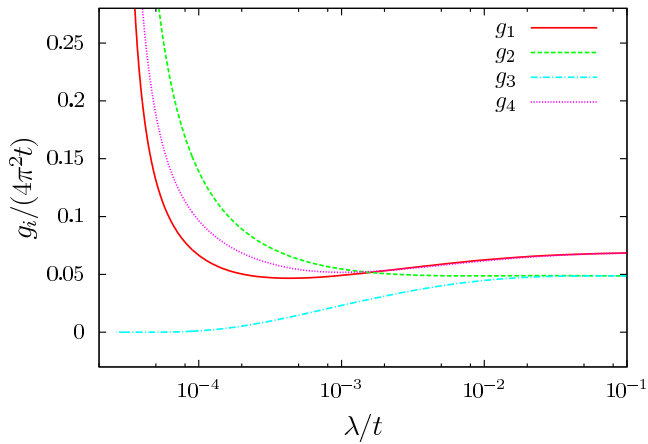}
\caption{(Color online) Flow without band-flip self-energy term for $ \Delta E = 6.5 $, $ w_d =11.38 $, $ w_s= 23.51$, $ U=0.03 \cdot 4 \pi^2 w_d $,
 $ U'=0.3 U $ at approximation level~1.
 Line styles as in Fig.~\ref{fig:nod-dSC}.}
 \label{fig:nod-FM}
\end{figure}

First we characterize the nature of the flow in the absence of a six-point term. Characteristic curves for the flow of the couplings $g_i$ for this approximation level~1 can be seen in Figs.~\ref{fig:nod-dSC} and~\ref{fig:nod-FM}.
For zero next-to-nearest neighbor hopping, the Fermi surface is perfectly nested, giving rise to antiferromagnetism as the leading instability.
If $t'$ is small but nonzero, the most important loops in the flow equations Eqs.~(\ref{eqn:gdot1}-\ref{eqn:gdot4}) are $ d_0 $ and $ d_1 $. Therefore $ d_2 $ and $ d_3 $ 
can be neglected at first.
In this approximation $ g_4 $ is 
decreased by $ d_0 $ while the $ d_1 $ term in Eq.~(\ref{eqn:gdot3}) prevents $ g_3 $ from being renormalized to zero. This allows for a sign change of $ g_4 $
after which the growth of $ d_0 $ drives the system to a $d$-wave superconducting $ d $SC instability corresponding to a strong coupling fixed point with $ g_4 \to - \infty $ and
 $ g_1,g_2,g_3 \to + \infty $. This instability is in full agreement with the findings of previous two-patch studies\cite{furukawa_1998}. 
 If we now increase $t'$, the particle-hole diagrams with small wavevector transfer become more important, i.e.\ the $ d_2 $ terms cannot be neglected any more. In particular, the last term in Eq.~(\ref{eqn:gdot4}) hampers the sign change of $ g_4 $, which now occurs at a lower scale, leading to a lower critical scale. Moreover, $ g_2 $ is now renormalized to zero instead of diverging to $ + \infty $. Thus we have an altered strong-coupling fixed point, but still with $ g_3\to + \infty$, $ g_4 \to - \infty$,  corresponding to a $ d $-wave pairing instability. This distinction was absent in traditional two-patch studies\cite{furukawa_1998,schulz} in which the small-wavevector-transfer particle-hole channel was disadvantaged by the choice of the cutoff function. 
 Katanin and Kampf have included such contributions in a momentum-shell approach\cite{Katanin-two-patch}. They find a similar strong coupling fixed point with non-diverging $ g_2 $ which however corresponds to an antiferromagnetic instability.
In $N$-patch flows we will a smooth crossover from the fixed point with $g_2 \to \infty$ to the one with $g_2 \to 0$.

\begin{figure}
 \includegraphics[width=8.5cm]{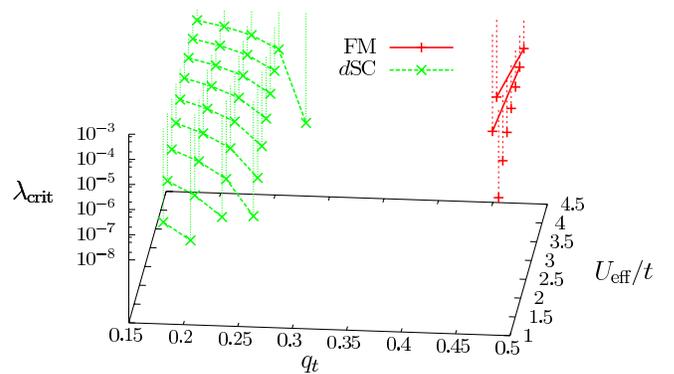}
 \caption{(Color online) Phase diagram of the $ t$-$ t' $ Hubbard model in two-patch approximation, approximation level~1, with $ q_t = - t' /t $.
 Points that correspond to the same value of $ U_{\rm eff} $ are connected by a line. The region around the separatrix between the $ d$SC and FM instabilities is inaccessible in our approach, since the numerical integration of the loops gets too cumbersome.}
 \label{fig:ob_hub}
\end{figure}

If the dispersion is varied further by increasing $t'$, $ d_2 $ grows even more strongly. Then it eventually prevents
 a sign change of $ g_4 $ and therefore excludes $ d $-wave superconductivity. Then the flow corresponds to a FM instability.
Since the $ d_2 $ terms in Eqs.~(\ref{eqn:gdot1}-\ref{eqn:gdot2}) depend linearly on $ g_4 $, its sign change to negative values occurring in the $d$-wave regime prohibited ferromagnetism, which reflects the mutual exclusion of the FM and $ d $SC instabilities. Along the separatrix between these two regimes, all four running couplings flow to zero. This suggests that the critical scale drops to zero from both sides, which implies the existence of a quantum critical
 line between the two phases. 
 In an $ N $-patch study, however, the situation is more involved and both instabilities may occur
 simultaneously. Unfortunately, the region around the separatrix is unaccessible in our calculations due to an excessive number of function calls required for numerical integration of the loops. 
In Fig.~\ref{fig:ob_hub}, we plot the phase diagram of this model obtained with approximation level~1.
In the $ d $-wave regime a reduction of the interaction strength to $ U_{\rm eff} $ leads to a lower critical scale.

Let us now turn to the case of non-vanishing inter-orbital coupling, where we have
$ g_1 = g_4 > g_2 = g_3 $ at the initial scale. From Table~\ref{tab:crit} we find that this detuning leads to a lower critical scale in the $ d $SC regime, as the inter-orbital interaction suppresses $ g_3 - g_4 $ in the initial condition while $ g_3 +g_4 $ remains unchanged.

Before we analyze the $ d $-wave regime in further detail, 
we briefly have a look at the band structure for parameters given for $ {\rm Y Ba_2 Cu_3 O_7} $ in Ref.~\onlinecite{andersen_jpcs_1995}. 
For such a system, however, van Hove filling is not close to the experimental situation, since it corresponds to a filling factor of about 0.19. 
The result should therefore not be taken as a realistic prediction for this material. In Fig.~\ref{fig:nod-FM} we observe that feedback the flow approaches a FM fixed point at level~1. This also holds for the other approximation levels.

 \begin{table}
 \begin{tabular}{ccccccc|c}
   $ \Delta E $ & $ w_d $ & $ w_s $ & $ n_{\rm low} $ & $ c_{\rm max}$ & $ t $& $ q_t $ & FIG.\\
\hline
 19.5 & 11.5 & 4.5 &  0.415 & 0.383 & 2.57 & 0.264 &\ref{fig:nod-dSC}\\
 6.5 & 11.38 & 23.51 & 0.192 & 0.503 & 3.61& 0.492 & \ref{fig:nod-FM}\\
 \end{tabular}
 \caption{Different parameter sets for the dispersion analyzed in this paper.
The lower-band filling factor $ \int_{\rm BZ} \! d {\bf k} \, \Theta [ - E_- ({\bf k}) ] / (4 \pi^2) $ is denoted by $ n_{\rm low} $ and the maximal hybridization matrix element by
$ c_{\rm max} = \max_k c_k $ .}
 \label{tab:param}
\end{table}

\begin{table}
 \begin{tabular}{l|cc|cc|cc}
   & \multicolumn{2}{c}{level~1} &\multicolumn{2}{c}{level~2} &\multicolumn{2}{c}{level~3} \\
   FIG. & $ \lambda_{\rm crit,1}/t $ & &$ \lambda_{\rm crit,2} /t $ & &$ \lambda_{\rm crit,3} /t $ & \\
\hline
  \ref{fig:nod-dSC}a) & $ 4.73 \cdot 10^{-6} $& $ d $SC & $1.28 \cdot 10^{-5} $ & $ d $SC &$ 1.32 \cdot 10^{-5}$ & $ d $SC\\
  \ref{fig:nod-dSC}b) & $ 3.83 \cdot 10^{-6} $& $ d $SC & $8.69 \cdot 10^{-6} $ & $ d $SC &$ 8.93 \cdot 10^{-6}$ & $ d $SC\\
  \ref{fig:nod-FM} & $ 3.76 \cdot 10^{-5} $& FM & $3.58 \cdot 10^{-5} $ & FM & $ 3.58 \cdot 10^{-5} $ & FM\\
 \hline
 \end{tabular}
 \caption{Critical scale $ \lambda_{\rm crit} $ and leading instability for different levels of approximation.}
 \label{tab:crit}
\end{table}

\subsection{Inclusion of the six-point term} \label{sec:num6pt}
\begin{figure*}
 \includegraphics[width=17.8cm]{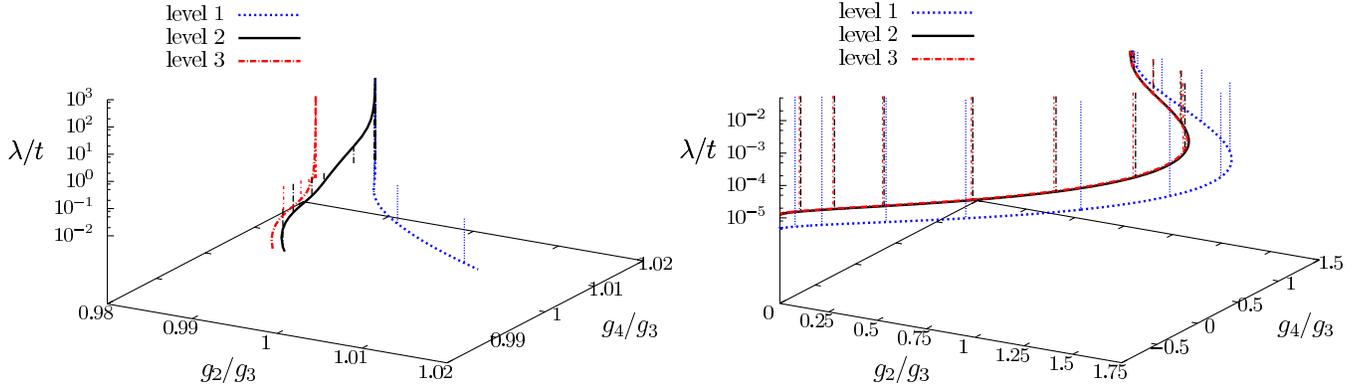}
 \caption{(Color online) Flows in the low-energy model for $ \Delta E =19.5 $, $ w_d=11.5 $, $w_s=4.5 $, $ U=0.03 \cdot 4 \pi^2 w_d $,
 $ U'=0 $ at the different approximation levels. The interaction has been projected to the $ g_2/g_3$-$g_4/g_3$ plane, while the $ z $-direction corresponds to the scale. The impulse-type vertical lines have been added for clearness and have no physical meaning. The initial flow for high scales is shown on the left hand side. Here one can see that approximation level 2 with six-point feedback has the same initial condition for the $g_i$s as level 1, but approaches quickly level~3 
in the early flow.  On the right hand side, the continuation of the flow at lower scales is depicted. Level 2 and 3 are basically equivalent and have a higher critical scale than level 1. The ratio $g_2/g_3$ flows to zero, corresponding to the second $d$-wave pairing fixed point discussed in Subsec.~\ref{sec:numwo}.
The band-flip self-energy term in the feedback has not been taken into account.}
 \label{fig:comp-nup}
\end{figure*}

\begin{figure*}
 \includegraphics[width=17.8cm]{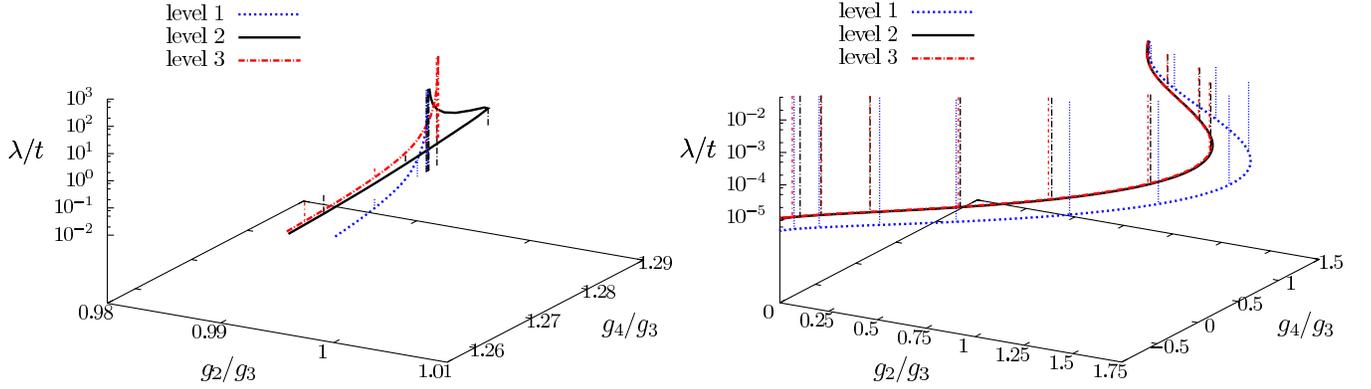}
 \caption{(Color online) Same as in Fig.~\ref{fig:comp-nup}, but for $ U'=0.3 U$.}
 \label{fig:comp-up}
\end{figure*}

 We now turn our attention to the six-point feedback for the band structure underlying the flows in Figs.~\ref{fig:comp-nup} and~\ref{fig:comp-up}, employing the improved approximation levels 2 and 3 introduced in Sec.~\ref{subsec:imptrunc}. Note that the model parameters will be varied in order to clarify the differences between these approximations and may not always correspond to an experimentally realistic situation.
 
\begin{figure}
  \includegraphics[width=8.5cm]{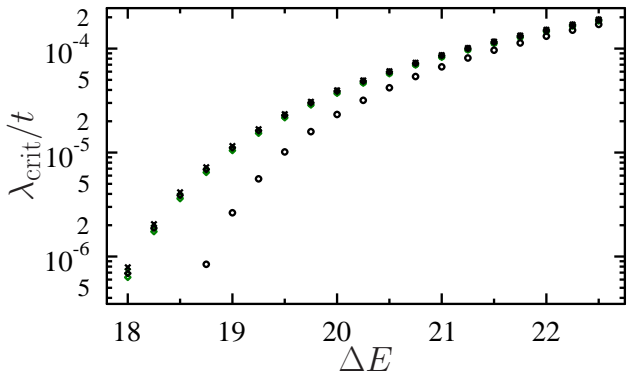}
 \caption{(Color online) Critical scale $ \lambda_{\rm crit} $ for  $ w_d=11.5 $, $w_s=4.5 $, $ U=0.03 \cdot 4 \pi^2 w_d $,
 $ U'=0.3 U $ as a function of $ \Delta E $, in the $d$SC regime. Circles correspond to approximation level~1, diamonds to level~2 and crosses to level~3.  Green and black markers represent values obtained with and without band-flip self-energies, respectively.}
 \label{fig:vardelt}
\end{figure}

We observe a flow to a $ d $SC strong-coupling fixed point over a wide parameter range for all approximation levels considered. This pairing instability occurs irrespective of the presence of inter-orbital interactions. The critical scale, however, is enhanced by the six-point feedback.
 Some numbers can be inferred from Table~\ref{tab:crit} and also Fig.~\ref{fig:vardelt}, the enhancement can easily be a factor of two, at least in this parameter range of small instability scales.
We observe that $ \lambda_{\rm crit} $ only differs weakly
between the approximation levels 2 and 3. Figs.~\ref{fig:comp-nup} and~\ref{fig:comp-up} illustrate that first the mixed-band diagrams flow to a value close to their infrared limit before the lower-band diagrams start to grow significantly. (All other ratios of the couplings that are not shown in Figs.~\ref{fig:comp-nup} and~\ref{fig:comp-up} behave indeed likewise.) This separation of scales is enhanced or might be even induced by a two-patch approximation. It ensures that the two-loop correction term discussed in Appendix~\ref{sec:two-loop} remains negligible and also holds in the ferromagnetic case and for Fig.~\ref{fig:vardelt}.

The question now is whether the six-point feedback on the critical scale can be related to characteristic properties of the band structure such as band curvatures.
For this purpose, we consider $ \Delta V_- $ terms in the initial condition of approximation level~3 for small hybridization $ c_k $ as in the case of Fig.~\ref{fig:comp-nup}. If for simplicity $ U' $ is then sent to zero,
 the bare coupling functions $ V_- $ and $ V_3 $ Eqs.~(\ref{eqn:bare-coup-})
and (\ref{eqn:bare-coup3}) read in leading order in the hybridization
\begin{align*}
 V_- (k_1,k_2,k_3) & \approx U \prod_i d_{k_i}  \\
  V_{3} (k_1,k_2,k_3) & \approx  - U  c_{k_1} d_{k_2} d_{k_3} d_{k_4} \, .
\end{align*}
 This corresponds to taking only the on-site interaction in the $ d $-orbital into account. Since $ U'=0 $, the self-energy insertion contributions to all
four running couplings take on the same value and the direct particle-hole diagrams (\ref{eqn:Rd}) vanish. 
In the diagrams in Eq.~(\ref{eqn:1ld}) the hybridization $ c_k $ appears only
inside the integrand whereas the external legs of momentum $ {\bf k} $ have a factor $ d_k $, which in contrast to $ c_k $ is invariant under a spatial
rotation by $ \pi/2 $. This implies that the particle-particle contributions to $ \Delta V_-(A,A,B) $ and $ \Delta V_- (A,A,A) $ are of equal size, 
whereas the crossed particle-hole diagrams give different contributions. The latter can be seen as follows: After calculating the Matsubara sum for $\lambda = 0 $,
 the integral in Eq.~(\ref{eqn:Rcr}) reads as
\begin{equation} \label{eqn:hyb-int}
 \int_{\rm BZ} \!\!\!\! d{\bf q} \, \,
  c_{{\bf l}+{\bf q}}^2 \, d_{\bf q}^2 \, \left[ E_+ ({\bf q}+{\bf l} ) - E_- ({\bf q})  \right]^{-1} \Theta \left( - E_- ({\bf q}) \right) \, ,
\end{equation}
with the Heaviside function $ \Theta (x) $.
For $ \Delta V_- (A,A,A) $, which renormalizes $ g_4$,  we have $ {\bf l} = 0 $ and for 
$ \Delta V_- (A,A,B) $, which renormalizes $ g_3$, the transfer momentum is $ {\bf l} = \hat{\boldsymbol \pi} $.
 Since both bands have a curvature with the same sign in our model and since $ E_+ $ is always positive,
the denominator in the integrand of Eq.~(\ref{eqn:hyb-int}) should take on smaller values for $ {\bf l} = 0 $ 
than for $ {\bf l} = \hat{\boldsymbol \pi} $ on a large phase space region centered around $ {\bf q} = 0 $.
One might therefore expect a suppression of $ g_3 - g_4 $.
 This argument, however, ignores the momentum
dependence of the orbital weight completely.
 Whilst being zero along the diagonals of the BZ, the hybridization matrix elements $ c_{\bf q} $ have their maximal value
 close to the saddle points of the lower band. This weakens the effect of the band curvature on $ g_3 - g_4$,
in particular if the hybridization shows plateau-like structures centered around the van Hove points
 as for the parameters underlying the flow in Fig.~\ref{fig:comp-nup}.  
At these points, however, the four-fold rotation symmetry of the dispersion
gives rise to identical denominators of the loop
 integrand in Eq.~(\ref{eqn:hyb-int}) for $ {\bf l} = 0 $ and $ {\bf l} = \hat{\boldsymbol \pi} $.
 Hence, the effect of band curvatures that one might expect ignoring orbital makeup effects should
 play a minor role in such a case.
 In contrast, the phase space weight imposed by the orbital makeup may lead to an enhancement of $ g_3 - g_4 $, even for weak hybridization 
as in the case of Fig.~\ref{fig:comp-nup}.
 For larger values of the hybridization, the situation is yet more involved, as terms with opposite signs compete.
So far, we have discussed the contributions of $ \Delta V_- $ to $ g_3 - g_4 $ at approximation level~3. We have found that they are quite sensitive to 
the orbital makeup.
 The question now is how they affect the critical scale:
On average, $ \Delta V_- $ suppresses the couplings $ g_i $ while the difference $ g_3 - g_4 $ may be enhanced or lowered depending on the 
orbital makeup. In general, smaller initial values of the couplings give rise to lower critical scales while a larger value of the $d$-wave coupling $ g_3 - g_4 $
 promotes
a sign change of $ g_4 $ at higher scales.
 So for enhanced $ d $-wave coupling, we have to deal with two counteracting tendencies and it is not clear a priori which one prevails, whereas
a suppression of $ \lambda_{\rm crit} $ is to be expected if $ g_3 - g_4 $ is lowered. 
In Fig.~\ref{fig:comp-up} they lead to a surprisingly
large enhancement of $ \lambda_{\rm crit} $. Indeed, $ g_3 -g_4 $ gets larger when the six-point feedback is taken into account, but if we
neglect contributions from $ d_2 $ and $ d_3 $ in the flow equations Eqs.~(\ref{eqn:gdot1}-\ref{eqn:gdot4}) the critical scale is virtually the same as
in the conventional truncation. Moreover, its value is increased by orders of magnitude without the $ d_2 $ and $ d_3 $ terms. This points out the importance of 
those terms and their interplay with the six-point feedback for the dispersion underlying Fig.~\ref{fig:comp-nup}.

In Fig.~\ref{fig:comp-up} $ g_3 - g_4 $ is suppressed by $ \Delta V_- $. Without the $ d_2 $ and $ d_3 $ terms, this would only change 
the critical scale by values below the level of accuracy,
while $ \lambda_{\rm crit} $ is significantly enhanced if those terms are taken into account. As for the flow to a ferromagnetic instability 
in Fig.~\ref{fig:nod-FM}, we find that the six-point term is of minor importance for the corresponding parameters.

Finally, we investigate the impact of the band separation $ \Delta E $ on the critical scale as depicted in Fig.~\ref{fig:vardelt}. Lower values of $ \Delta E $
correspond to a larger ratio $ - t'/ t $ and therefore to a lower critical scale in the $ d $SC regime. We observe that for band separations smaller than $ 19.5 $
 units, the six-point feedback substantially enhances the critical scale. In particular, when the critical scale gets small due to the competition with the FM channel, the six-point term can change the result by an order of magnitude. This indicates that these corrections may play a role in situations with competing ordering tendencies. Deep in the $d$-wave regime, or also on the FM side, the impact of the six-point term is only of quantitative nature.
 Since the behavior at small $d$SC critical scales mainly stems from the interplay
of the six-point feedback and the $ d_2 $ loop, it should result from the increasing contribution of $ d_2 $ that we have to encounter 
when $ \Delta E $ is lowered.
 The inclusion of the band-flip self-energy term only leads to slight changes of
the critical scale and does not affect our result on a qualitative level.
\section{Conclusions}

In summary, we have proposed an fRG scheme for multiband systems that takes virtual excitations outside a low-energy window into account. This approach in a way extends the usual RG approach of accessing the low-energy physics of a given system, where the effective action is truncated after the four-point interaction. The conventional treatment completely neglects virtual excitations involving both high- and low-energy modes once one has switched to the low-energy theory.
 Our approach, in contrast, considers them up to second order in perturbation theory within a frame with a larger energy scale around such a low-energy window. On the formal level, this is achieved by a one-loop fRG treatment of an effective action for the low-energy modes, which is truncated {\em after} (i.e.\ keeps) the three-particle interactions generated by the high-energy modes (six-point term). Of course the low-energy theory could also be solved using a different method than fRG, and also here the six-point term of the effective interactions may play a role.

In the RG flow in the low-energy window, the six-point vertex gives rise to mixed one-loop diagrams with one low-energy and one high-energy leg. Actually, a part of these diagrams is summed up in the cRPA framework currently used  in ab-initio calculations\cite{aryasetiawan,imada}.  The correlation functions of high-energy modes are no longer kept track of in our scheme. This drastically reduces the required numerical  resources compared to a 'full' fRG approach with an extended energy window. 
 
We have numerically investigated the impact of the mixed diagrams that are absent in the conventional approach for a simple two-band model within the  two-patch approximation.
Leaving a large part of the Fermi surface unaccounted, this approximation is mainly chosen for reasons of technical simplicity. On the other hand, the two-patch treatment may appear questionable if virtual excitations in the upper band away from the Fermi level are considered, since renormalization effects at intermediate scales are discarded. However, the impact of the rest of the Fermi surface has been investigated for the one-band Hubbard model within $ N $-patch schemes\cite{metznerRMP}, and recently, within a channel decomposition\cite{husemann_2009}. The leading tendencies in the parameter region of interest are well captured in the two-patch treatment. So our results may indicate well how the virtual high-energy excitation affect the RG flow, also if the full Fermi surface was considered. A thorough fRG study of multiband models that takes into account the off-Fermi-surface bands perturbatively would require a multi-patch scheme that goes beyond Fermi-surface projection of the interactions. The authors are currently implementing such a scheme.
 
In this work, we have used a multiplicative frequency cutoff\cite{husemann_2009}, which allows for a second $ d$SC and a FM fixed point that cannot be found in two-patch studies using a traditional momentum-shell cutoff. This is due to the particle-hole diagrams with vanishing momentum transfer that contribute to the flow at nonzero scales if a multiplicative frequency cutoff is used. The regularization scheme of momentum-shell type put forward in Ref.~\onlinecite{Katanin-two-patch} also accounts
for these diagrams and gives results that are qualitatively similar to ours at approximation level~1.
However, it cannot be extended to a region outside the patches and is therefore not viable at approximation level~2.

In the two-patch treatment the approximation levels we call 2 and 3 merely coincide, that is the six-point term can be described by adding the mixed diagrams to the bare values of the couplings in the initial conditions of the flow. This represents a technically appealing approximation that can be implemented more easily compared to keeping a flowing six-point term. It should hold in cases where the mixed diagrams are renormalized at scales at which the low-energy diagrams only flow weakly.
Close to a separatrix in the flow between $d$SC and FM ordering, even slight changes in the parameters of the bare action, i.e.\ the initial conditions of the flow, can result in a drastic change of the critical scale. In the two-patch model, this can be understood by the small number of fixed points with distinct physical properties, which can be mutually exclusive. 
In an $ N $-patch approach this should in principle no longer hold, since there could be a larger number of fixed points that decay into (partly overlapping) classes sharing similar physical properties. Yet, the flows for the full Fermi surface still clearly show a close competition and a possible quantum critical point between $d$SC and FM instabilities.
Indeed, the sensitivity with respect to the initial conditions is strongly reflected in our study of different approximation levels to the effective theory. We find that the six-point feedback can enhance the critical scale by orders of magnitude in the $d$SC regime, provided that the particle-hole bubble with 
zero transfer momentum diverges strongly enough in the infrared, i.e.\ there is a strong competition between $d$SC and FM tendencies. Away from this critical region and on the FM side, the impact of the six-point term is reduced.  

Currently it is unclear if these differences will persist or get weakened if a multi-patch scheme is used, and if
the six-point feedback may still play an important role at least on a quantitative level in such a flow. Very recent results on two- and three-band models aiming at different trends in cuprate systems using related approximations gave a slight enhancement up to 15$\%$ due to the bands away from the Fermi level\cite{uebelacker}. 
Regarding the iron pnictide superconductors, there a several very interesting question that arise from this study. There it might be useful to carefully study
 other observables than just the critical scale.  For example, most RPA and fRG studies indicate that the orbital character of the bands induces a pronounced anisotropy of the sign-changing $s$-wave gap function around the various Fermi pockets. Of course, for the predictive power of these calculations it is necessary to check how these anisotropies are affected by the higher-order terms in the effective interactions.  Then, depending on the parameters of the multiband model, there is the possibility of a competition of other superconducting states with the anisotropic $s$-wave state. Again, in such situations additional effects may play a decisive role, and may either increase or decrease the degree of competition.

In summary, we have provided tractable approximations for improved effective interactions within the conduction band of a multiband problem and performed a case study in which the correction terms play a visible role. The next steps should include more extended applications in multiband models for correlated electron systems, in order to understand the importance of these corrections in a comprehensive way. 

\section*{Acknowledgments}

We thank Manfred Salmhofer, Stefan Uebelacker, Kay-Uwe Giering, and Jutta Ortloff for discussions. This work was supported by the DFG priority program SPP1458 on iron pnictide superconductors and by the DFG research unit FOR 723 on functional renormalization group methods. 

\begin{appendix}

\section{Parameterization of the interaction}\label{sec:paramapp}

We will now parameterize the interaction in a way similar to \cite{salm_hon_2001}. The quartic part of the action is therefore cast in a suitable form:
\begin{align*}
 S^{(4)} = & \,  \frac{U}{4}  \sum_{\alpha,\{k_i,\sigma_i\}}\left(E_{\boldsymbol \sigma} - D_{\boldsymbol \sigma} \right) \delta_k \\ 
& \times \bar{\psi}_{k_1,\alpha,\sigma_1} \bar{\psi}_{k_2,\alpha,\sigma_2}  \psi_{k_3,\alpha,\sigma_3}   \psi_{k_4,\alpha,\sigma_4}  \\
  + & \, \frac{U'}{4}  \sum_{\{\alpha_i,k_i,\sigma_i\}} \delta_{\alpha_1,-\alpha_2} 
\left( E_{\boldsymbol \sigma} E_{\boldsymbol \alpha} - D_{\boldsymbol \sigma} D_{\boldsymbol \alpha} \right)  \delta_k \\
&  \times \bar{\psi}_{k_1,\alpha_1,\sigma_1} \bar{\psi}_{k_2,\alpha_2,\sigma_2}  \psi_{k_3,\alpha_3,\sigma_3}   \psi_{k_4,\alpha_4,\sigma_4}    \, ,
\end{align*}
where $ D_{\boldsymbol \sigma} = \delta_{\sigma_1,\sigma_4} \delta_{\sigma_2,\sigma_3} $,
 $ E_{\boldsymbol \sigma} = \delta_{\sigma_1,\sigma_3} \delta_{\sigma_2,\sigma_4} $, and where $ \delta_k = \delta \left( k_1+k_2-k_3-k_4 \right) $ ensures energy
and momentum conservation.

Upon the orbital-to-band transformation the $ \delta $-function in the momenta gets multiplied by a product of form factors $ d_k $ and/or $ c_k $. 
We now rewrite $ S^{(4)} $ in terms of the fields $ \chi $ 
\begin{align*}
 S^{(4)} [\bar{\chi},\chi] = & \,  \frac{1}{4} \sum_{\{X_i\}} F(X_1,X_2,X_3,X_4) \\
& \times \bar{\chi}(X_1) \bar{\chi}(X_2) \chi(X_3) \chi (X_4)
\end{align*}
and decompose the interaction according to the band indices of the external legs. This yields
\begin{widetext}
\begin{align*}
F(X_1,X_2,X_3,X_4)  =  & \, {\cal A}_{X_1,X_2} {\cal A}_{X_3,X_4} \left[ F_+ (\xi_1,\xi_2,\xi_3,\xi_4) \, 
 \delta_{{\boldsymbol \alpha},++++} 
+ F_- (\xi_1,\xi_2,\xi_3,\xi_4) \, \delta_{{\boldsymbol \alpha},----} \right. \\ &
 + F_{+\to-} (\xi_1,\xi_2,\xi_3,\xi_4) \, \delta_{{\boldsymbol \alpha},--++} 
 + F_{-\to+} (\xi_1,\xi_2,\xi_3,\xi_4) \, \delta_{{\boldsymbol \alpha},++--} 
+ 4 F_{\pm\to\pm} (\xi_1,\xi_2,\xi_3,\xi_4) \, \delta_{{\boldsymbol \alpha},+--+}   \\ &
  + 2 F_{+\to\pm} (\xi_1,\xi_2,\xi_3,\xi_4) \, \delta_{{\boldsymbol \alpha},-+++}
+ 2 F_{-\to\pm} (\xi_1,\xi_2,\xi_3,\xi_4) \, \delta_{{\boldsymbol \alpha},+---}   \\ &   \left.
+ 2 F_{\pm\to+} (\xi_1,\xi_2,\xi_3,\xi_4) \, \delta_{{\boldsymbol \alpha},+++-} 
+ 2 F_{\pm\to-} (\xi_1,\xi_2,\xi_3,\xi_4) \, \delta_{{\boldsymbol \alpha},---+} \right] \, ,
\end{align*}
\end{widetext}
where $ \xi_i $ denotes $ (k_i,\sigma_i) $ and $ X_i = (\alpha_i,\xi_i) $ and where the two-point antisymmetrization operator $ {\cal A} $ has been defined
as $ {\cal A}_{a,b} f(a,b)= \left[ f(a,b)-f(b,a) \right] / 2 $.  
In the orbital picture, the interaction is invariant under a band-index flip. Since the trace of the matrix of the orbital-to-band transformation Eq.~(\ref{eqn:otb}) vanishes, this property also holds in the band language, giving rise to the following identities
\begin{align*}
 F_+ (\xi_1,\xi_2,\xi_3,\xi_4) & =  F_- (\xi_1,\xi_2,\xi_3,\xi_4) \\
 F_{+\to-} (\xi_1,\xi_2,\xi_3,\xi_4) & =  F_{-\to+} (\xi_1,\xi_2,\xi_3,\xi_4) \\
 F_{\pm\to\pm} (\xi_1,\xi_2,\xi_3,\xi_4) & =  F_{\pm\to\pm} (\xi_2,\xi_1,\xi_4,\xi_3) \\
 F_{-\to\pm} (\xi_1,\xi_2,\xi_3,\xi_4) & =  F_{+\to\pm} (\xi_1,\xi_2,\xi_3,\xi_4) \\
 F_{\pm\to-} (\xi_1,\xi_2,\xi_3,\xi_4) & =  F_{\pm\to+} (\xi_1,\xi_2,\xi_3,\xi_4) \, .
\end{align*}
Hermiticity of the Hamiltonian in the band language requires that the band conserving terms $ F_-$, $ F_+$, $F_{\pm\to\pm} $ must obey
\begin{equation}  \label{eqn:conj}
  F_{X} (\xi_1,\xi_2,\xi_3,\xi_4) = F_{X}^\ast (\xi_4,\xi_3,\xi_2,\xi_1) \, .
\end{equation}
 For the non-band-conserving terms it leads to the following relations:
\begin{align*}
 F_{2} (\xi_1,\xi_2,\xi_3,\xi_4) : &  = F_{+\to-} (\xi_1,\xi_2,\xi_3,\xi_4) \\
&= F_{-\to+}^\ast (\xi_4,\xi_3,\xi_2,\xi_1) \\
 F_{3} (\xi_1,\xi_2,\xi_3,\xi_4) : &  = F_{-\to\pm} (\xi_1,\xi_2,\xi_3,\xi_4)\\
& = F_{\pm\to-}^\ast (\xi_4,\xi_3,\xi_2,\xi_1) \, .
\end{align*}
The anticommuting nature of the Grassmann fields is reflected by the following antisymmetry constraint
\begin{align*}
 F_X (\xi_1,\xi_2,\xi_3,\xi_4)  & =  - F_X (\xi_2,\xi_1,\xi_3,\xi_4) \\
&= - F_{X} (\xi_1,\xi_2,\xi_4,\xi_3) \, ,
\end{align*} 
for  $ X=+,-,2 $. According to \cite{salm_hon_2001}, these U(1)-vertices can be parameterized as
\begin{align*}
 F_{X} (\xi_1,\xi_2,\xi_3,\xi_4) = \,  & \delta_k\, \left[  E_{\boldsymbol \sigma} V_{X} (k_2,k_1,k_3) \right. \\
 & \left.  - D_{\boldsymbol \sigma} V_{X} (k_1,k_2,k_3) \right] \, ,
\end{align*}
where  $ V_X (k_1,k_2,k_3) = V_X (k_2,k_1,k_3-k_1-k_2) $.
For vertices with three legs on one band and one on the other, however, one of the symmetry constraints is violated and we only have
$ F_3 (\xi_1,\xi_2,\xi_3,\xi_4) = - F_3 (\xi_1,\xi_2,\xi_4,\xi_3) $,
which allows as well for the parameterization
\begin{align*}
 F_{3} (\xi_1,\xi_2,\xi_3,\xi_4) = &  \,\delta_k\, \left[  E_{\boldsymbol \sigma} V_{3} (k_1,k_2,k_4) \right. \\
& \left. - D_{\boldsymbol \sigma} V_{3} (k_1,k_2,k_3) \right] \, .
\end{align*} 
In contrast to the vertices  $ V_X $ considered before, there is no symmetry constraint for $ V_3 $. Finally, there are no antisymmetry relations for $F_{\pm\to\pm}$, which gives rise to the parameterization
\begin{align*}
 F_{\pm\to\pm} (\xi_1,\xi_2,\xi_3,\xi_4) = & \, \delta_k\, \left[  E_{\boldsymbol \sigma} V^{(E)}_{\pm\to\pm} (k_1,k_2,k_3) \right. \\
& \left. - D_{\boldsymbol \sigma} V^{(D)}_{\pm\to\pm} (k_1,k_2,k_3) \right] \, ,
\end{align*}
with  the symmetry constraint reflecting the behavior under a band-index flip. 
$ V^{(D,E)}_{\pm\to\pm} (k_1,k_2,k_3) = V^{(D,E)}_{\pm\to\pm} (k_2,k_1,k_3-k_1-k_2) $.
We now give explicit expressions for the coupling functions
\begin{widetext}
\begin{align} \label{eqn:bare-coup-}
 V_- (k_1,k_2,k_3) & = U \left( \prod_i d_{k_i} + \prod_i c_{k_i} \right) 
+ U' \left( d_{k_1} c_{k_2} c_{k_3} d_{k_4} + c_{k_1} d_{k_2} d_{k_3} c_{k_4} \right) \\ \notag
V_{2} (k_1,k_2,k_3) & =  U \left( d_{k_1} d_{k_2} c_{k_3} c_{k_4} + c_{k_1} c_{k_2} d_{k_3} d_{k_4} \right) -
U' \left( d_{k_1} c_{k_2} d_{k_3} c_{k_4} + c_{k_1} d_{k_2} c_{k_3} d_{k_4} \right) \\ \notag
  V_{\pm\to\pm}^{(D)} (k_1,k_2,k_3) & = U\left( d_{k_1} c_{k_2} c_{k_3} d_{k_4} + c_{k_1} d_{k_2} d_{k_3} c_{k_4} \right) +
  U' \left( \prod_i d_{k_i} + \prod_i c_{k_i} \right) \\ \notag
  V_{\pm\to\pm}^{(E)} (k_1,k_2,k_3) & =  U\left( d_{k_1} c_{k_2} c_{k_3} d_{k_4} + c_{k_1} d_{k_2} d_{k_3} c_{k_4} \right) -
  U' \left(  c_{k_1} c_{k_2} d_{k_3} d_{k_4} + d_{k_1} d_{k_2} c_{k_3} c_{k_4}\right) \\ \label{eqn:bare-coup3}
  V_{3} (k_1,k_2,k_3) & =  U \left(  d_{k_1} c_{k_2} c_{k_3} c_{k_4} - c_{k_1} d_{k_2} d_{k_3} d_{k_4} \right) +
 U' \left( d_{k_1} d_{k_2} d_{k_3} c_{k_4} - c_{k_1} c_{k_2} c_{k_3} d_{k_4}  \right) \, .
\end{align}
\end{widetext}
  
 So the bare interaction can be expressed in terms of 5 independent functions of three $ 1+2$-momenta by exploiting its symmetries.
\section{Flow equations} \label{sec:6ptapp}

 In this appendix, the flow equations used in this paper shall be given explicitly. We neglect self-energy effects and put the six-point
 vertex to be constant during the flow. Therefore, only the four-point flow equation matters. If the six-point function is ignored, 
the right hand side of this equation is given by Eq.~(88) in Ref.~\onlinecite{salm_hon_2001} with $ V_- $ as an initial condition for $ V $. 
In this form, the flow equation can be used to discuss the impact of orbital makeup.
Proceeding further, we take into account the feedback term
\begin{align*}
 \Delta_4 (\xi_1,\xi_2,\xi_3,\xi_4) = & \, - \frac{1}{2} \int \! d{\boldsymbol \eta}  \,  S(\eta_1,\eta_2) \\
& \times F^{(6)} (\eta_1,\xi_1,\xi_2,\eta_2,\xi_3,\xi_4) \\
 \Delta_4 (\xi_1,\xi_2,\xi_3,\xi_4) = & \, \delta_k\, \left[  E_{\boldsymbol \sigma} \, \delta V_- (k_2,k_1,k_3)  \right. \\
& \left. - D_{\boldsymbol \sigma} \, \delta V_- (k_1,k_2,k_3) \right] \, ,
\end{align*}
$ S $ denoting the single scale propagator $ {\boldsymbol S} = \dot{\boldsymbol G} - {\boldsymbol G} \dot{\boldsymbol \Sigma} {\boldsymbol G}$ with
  self-energy $ \Sigma $. In the following, we assume the coupling functions $ V_- $ and $ V_3 $ to be real as is the case for the model analyzed in this work.
Up to second order in the bare interaction, the six-point coupling function is given by
\begin{align} \notag
 F^{(6)} ( \xi_1 \dots \xi_6) & = - 9 \, {\cal A}_{\xi_1,\xi_2,\xi_3} {\cal A}_{\xi_4,\xi_5,\xi_6}  \int \! d \boldsymbol{\eta} \, G_+ (\eta_1,\eta_2) \\ \label{eqn:6pt}
& \times F_3 (\eta_1,\xi_2,\xi_4,\xi_5) \, F_3 (\eta_2,\xi_6,\xi_3,\xi_1) \, \, ,
\end{align}
where $ G_+ $ denotes the upper-band propagator and where the three-point antisymmetrization operator
\begin{equation*}
 {\cal A}_{a,b,c} f(a,b,c) = \frac{1}{3!} \sum_\pi P_\pi \, f\left(\pi(a),\pi(b),\pi(c) \right).
\end{equation*}
 is given by the difference of the sums of cyclic ($ P_\pi =1 $) and anti-cyclic ($ P_\pi =-1 $) permutations $ \pi $.
These antisymmetrization operators give rise to self-energy insertion and one-loop diagrams
 $ \Delta_4  = \Delta_{\rm SE} + \Delta_{\rm loop} $. 
The former  contain the band-flip self-energy
\begin{equation*} 
\Sigma_{\pm} (\xi_1,\xi_2) = \delta\! (\xi_1-\xi_2) \int \!\! d\eta_1 \, d\eta_2 \, F_3 (\xi_1,\eta_1,\eta_2,\xi_2) \, S(\eta_1,\eta_2)
\end{equation*}
and read as
\begin{align*}
 \Delta_{\rm SE} (\xi_1,\xi_2,\xi_3,\xi_4) = & \, \int \! d\eta_3 \, d \eta_4 \, G_+ (\eta_3,\eta_4)\\
 \times & \, \left[  {\cal A}_{\xi_1, \xi_2} \, F_3 (\eta_3,\xi_2,\xi_3,\xi_4)\,  \Sigma_\pm (\eta_4,\xi_1) \right. \\
     + & \, \left. {\cal A}_{\xi_3, \xi_4} \, F_3 (\eta_4,\xi_3,\xi_2,\xi_1)\,  \Sigma_\pm (\eta_3,\xi_4)   \right] \, .
\end{align*}
This expression can again be parameterized as
\begin{align} \notag
 \Delta_{\rm SE} (\xi_1,\xi_2,\xi_3,\xi_4) = & \, \delta_k\, \left[  E_{\boldsymbol \sigma} \, V_{\rm SE} (k_2,k_1,k_3)  \right. \\
\notag
& \left. - D_{\boldsymbol \sigma} \, V_{\rm SE} (k_1,k_2,k_3) \right] \, ,
\end{align}
where $ V_{\rm SE} $ obeys the same symmetry constraints as $ V_- $.
This gives rise to 
\begin{align*}
 V_{\rm SE} (k_1,k_2,k_3) & =  V_3 (k_1,k_2,k_3) \tilde{\Sigma}_\pm (k_1) \\ & +
V_3 (k_2,k_1,k_1+k_2-k_3) \tilde{\Sigma}_\pm (k_2)  \\ & +
V_3 (k_1+k_2-k_3,k_3,k_2) \tilde{\Sigma}_\pm (k_1+k_2-k_3) \\ & +
V_3 (k_3,k_1+k_2-k_3,k_1) \tilde{\Sigma}_\pm (k_3) 
\end{align*}
 with 
\begin{equation*}
 \tilde{\Sigma}_\pm (k) = G_+ (k) \int dq \, S(q) \left[ V_3 (k,q,k) - 2 V_3 (k,q,q) \right]
\end{equation*}
The one-loop part comprises particle-particle, crossed and direct particle-hole diagrams
\begin{align} \notag
 \Delta_{\rm loop} (\xi_1,\xi_2,\xi_3,\xi_4) = & \, \delta_k\, \left[  E_{\boldsymbol \sigma} \, V_{\rm loop} (k_2,k_1,k_3)  \right. \\
\notag
& \left. - D_{\boldsymbol \sigma} \, V_{\rm loop} (k_1,k_2,k_3) \right] \\ \notag
  V_{\rm loop} (k_1,k_2,k_3)  = & \, {\cal R}_{\rm pp} (k_1,k_2,k_3) + {\cal R}_{\rm ph,cr} (k_1,k_2,k_3) \\ \label{eqn:1ld}
&  + {\cal R}_{\rm ph,d} (k_1,k_2,k_3)  \, .
 \end{align}
Since $ V_3 $ obeys no symmetry constraint, the particle-particle contribution
\begin{widetext}
\begin{align*}
  {\cal R}_{\rm pp} (k_1,k_2,k_3)  = - \int\! dq  \, S(q) \, G_+(l-q) & \left[ V_3 (l-q,q,k_1) V_3 (l-q,q,l-k_3) \right. \\
 +  & \left.  V_3 (l-q,q,k_2) V_3 (l-q,q,k_3) \right]_{l=k_1+k_2}
\end{align*}
consists of two terms that otherwise would coincide. The crossed particle-hole terms
\begin{align} \notag
  {\cal R}_{\rm ph,cr} (k_1,k_2,k_3)   = & - \int\! dq  \, S(q) \, G_+(l+q)  \left. V_3 (l+q,k_1,q) V_3 (l+q,k_2-l,q) \right|_{l=k_3-k_1} \\
 \label{eqn:Rcr}
   & - \int\! dq  \, S(q) \, G_+(l+q)  \left. V_3 (l+q,k_2,q) V_3 (l+q,k_3,q) \right|_{l=k_1-k_3}
\end{align} 
behave likewise. The direct particle-hole diagrams read as
\begin{align} \notag
  {\cal R}_{\rm ph,d} & (k_1,k_2,k_3) = - \int\! dq \,  S(q) \, G_+(l+q)  \left[ -2 V_3 (l+q,k_1,k_1+l) V_3 (l+q,k_3,k_2) \right. \\ \notag
 & +  \left.  V_3 (l+q,k_1,k_1+l) V_3 (l+q,k_3,q) + V_3 (l+q,k_1,q) V_3 (l+q,k_3,k_2) \right]_{l=k_2-k_3} \\ \notag
 - & \int\! dq \,  S(q) \, G_+(l+q)  \left[ -2 V_3 (l+q,k_2,k_3) V_3 (l+q,k_1-l,k_1) \right. \\ \label{eqn:Rd}
 & +  \left.  V_3 (l+q,k_2,k_1-l) V_3 (l+q,k_3,q) + V_3 (l+q,k_2,q) V_3 (l+q,k_1-l,k_1) \right]_{l=k_3-k_2} \, .
\end{align} 
\end{widetext}
 Note that for vanishing inter-orbital interaction $ U' =0 $ the direct particle-hole diagrams vanish due to $ V_3 (k_1,k_2,k_3) = V_3 (k_1,k_2,k_1+k_2-k_3) $.

\section{Two-loop corrections} \label{sec:two-loop}
\begin{figure}
 \includegraphics[width=8.5cm]{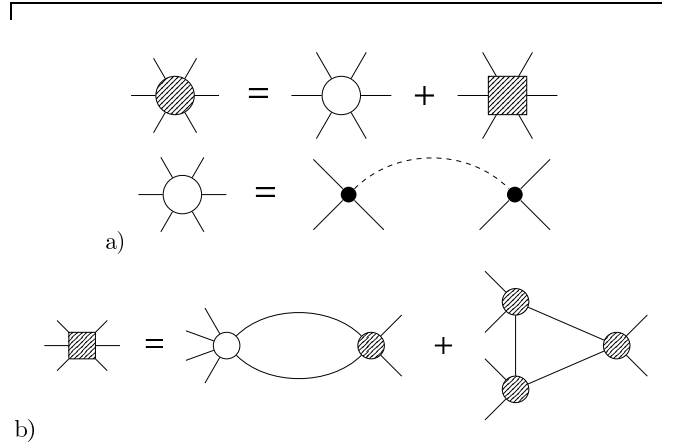}
 \caption{a) Full (scale dependent) vertex (hatched circle) written as a sum of its initial value (empty circle) and a renormalization induced correction term (hatched square).
b) Leading order result for the correction term.}
\label{fig:corr-6pt}
\end{figure}

Here, we give an argument why not going beyond level~2, i.e.\ why dropping the renormalization of the six-point vertex itself, may suffice.
 If one had not done so, a correction 
term would have to be added to the constant three-particle vertex as in Fig.~\ref{fig:corr-6pt}a). 
By integrating and then iterating the flow equation for the six-point vertex, this renormalization induced term can be expressed as a sum of diagrams containing
 the four-point and constant six-point vertices only up to arbitrary order. Note that the constant part of the six-point vertex is of second order in the four-point 
couplings.
In leading, i.e.\ third order, we obtain two one-loop terms depicted in Fig.~\ref{fig:corr-6pt}b). One of these diagrams includes only four-point vertices and would 
as well be present in the absence of an initial six-point term while the other one contains this initial three-particle interaction. When the right-hand side
of Fig.~\ref{fig:corr-6pt}b) is now fed
back into the flow equation for the four-point vertex, diagrams with overlapping and non-overlapping loops arise. If we had started our fRG analysis
directly from the full action $ \left( S^{(2)} + S^{(4)} \right) [\bar{\chi},\chi] $ instead of the effective low-energy action $ S_{\rm eff} $ and kept the band indices
as variables attached to the legs of the vertices, the overlapping ones would be neglected in the Katanin truncation. So they shall be dropped in the present
treatment as well. 

\begin{figure}
 \begin{center}
 \includegraphics[width=8.5cm]{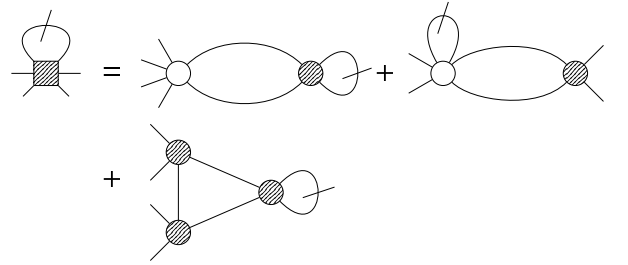}
 \end{center}
 \caption{Leading corrections to flow equation in Fig.~\ref{fig:feedback}. Diagrams with overlapping loops have been neglected.}
\label{fig:corr-flow}
\end{figure}

We then end up with the correction terms depicted in Fig.~\ref{fig:corr-flow}. The first and the third one can be merged with the one-loop terms
in Fig.~\ref{fig:feedback} leading to a Katanin substitution $ {\boldsymbol S} \to \dot{\boldsymbol G} $ both in the low-energy term as in the one-loop feedback.
 Since we neglect self-energy effects in our numerics, this substitution will not change our results for the six-point feedback.

The second diagram in Fig.~\ref{fig:corr-flow}, however, requires more care. We now proceed with giving upper estimates for the remaining correction term 
and the one-loop feedback and low-energy terms in Fig.~\ref{fig:feedback}. 
If the frequency dependence of the coupling functions is dropped, all Matsubara sums can
be evaluated analytically giving rise to the following rules for an estimate. 
\begin{itemize}
\item The four-point coupling functions $ V_3 $ and $ V_- $ at scale $ \lambda $ are replaced by their maximal value
$ g_3 $ and $ g_- $, respectively.
\item  Mixed loops including a scale derivative are replaced by a factor $ 4 \pi^2 ( \epsilon_+ \lambda)^{-1} $, where $ \epsilon_+ $
denotes the minimal energy of the high-energy bands. The band-flip self-energy diagrams behave likewise.
\item Lower-band loops with and without a scale derivative are replaced by $ 4 \pi^2 \lambda^{-2} $ and $ 4 \pi^2 \lambda^{-1} $, respectively.
\end{itemize}
At all scales the correction term Fig.~\ref{fig:corr-flow} should be small compared to the the low-energy term, which implies 
\begin{equation} \label{eqn:trunc-cond}
 4 \pi^2 g_3^2 \ll g_- \epsilon_+ \, . 
\end{equation}
Note that the orbital makeup reduces the value of the bare coupling functions and may therefore finally allow for the
omission of the correction term.
At scales at which the one-loop feedback term flows, the one-loop feedback should prevail against two-loop corrections, i.e.\ $ 4 \pi^2 g_- \ll \lambda $.
Together with the condition Eq.~(\ref{eqn:trunc-cond}), this requires the cutoff $ \lambda $ to be much larger than $ g_3^2 / \epsilon_+ $.
At lower scales
$ \lambda \ll g_-^2 \epsilon_+ /g_3^2 $, however, the mixed one-loop diagrams eventually saturate and the feedback term becomes negligible compared to the low-energy loop term that may finally drive the flow to a strong coupling fixed point.
 The crossover region between these two regimes should be small, as long as the inequality~(\ref{eqn:trunc-cond}) holds.
So a one-loop fRG approach should suffice to qualitatively discuss the impact of the six-point term on the critical scale for $ d $-wave superconductivity, for example.

Finally, we feel that a comment on the relation of the flows of $ \Gamma $ and $ \Gamma_- $ is in order. If self-energy effects are neglected
completely, the four-point vertex of $ \Gamma_- $ is equal to the four-point vertex of $ \Gamma $ with all external legs on the low-energy bands, since they both must
lead to the same four-point correlation functions for the low-energy modes. If one now considers the flow of $ \Gamma $ for an energy shell cutoff in the usual truncation and in addition forces
all four-point vertices with at least one leg on the high-energy bands not to flow, the one-loop feedback in the flow of $ \Gamma_- $ is recovered.
In the RG flow of $ \Gamma $, the most relevant correction term to this approximation consists of the diagram with the internal lines on the low-energy bands
and  three external legs on the low and one on the high-energy bands. If this correction is fed back into the flow of the vertex with all external legs on
the lower bands, the feedback correction term is equivalent to the second diagram in Fig.~\ref{fig:corr-flow} in leading order.
 As soon as self-energy effects are taken into account, however, this correspondence breaks down.

\end{appendix}

\bibliography{biblio}

\end{document}